\documentclass[openacc]{rstransa}%%%%where rstrans is the template name
\newcommand*{\doi}[1]{\href{http://dx.doi.org/#1}{doi: #1}}
\begin{document}

%%%% Article title to be placed here
\title{The solution of the Sixth Hilbert Problem: the Ultimate Galilean Revolution}

\author{%%%% Author details
Giacomo Mauro D'Ariano$^{1,2}$}

%%%%%%%%% Insert author address here
\address{$^{1}$Dipartimento di Fisica, Universit\`a degli Studi di Pavia, via Bassi 6, 27100 Pavia,\\
$^{2}$ INFN, Gruppo IV, Sezione di Pavia}

%%%% Subject entries to be placed here %%%%
\subject{}

%%%% Keyword entries to be placed here %%%%

%%%% Insert corresponding author and its email address}
\corres{Giacomo Mauro D'Ariano\\
\email{dariano@unipv.it}}

%%%% Abstract text to be placed here %%%%%%%%%%%%
\begin{abstract}
%Wigner called the effectiveness of mathematics in physics "unreasonable". Against such widespread romantic position 
I argue for a full mathematisation of  the physical theory, including its axioms, which must contain no physical primitives. In provocative words: ``physics from no  physics''. Although this may seem an oxymoron, it is the royal road to keep complete logical coherence, hence  falsifiability of the theory. For such a purely mathematical theory the physical connotation must pertain only the interpretation of the mathematics,  ranging from the axioms to the final theorems. On the contrary, the postulates of the two current major physical theories either don't have physical interpretation (as for von Neumann's axioms for quantum theory), or contain physical primitives as ``clock'', ``rigid rod '', ``force'',  ``inertial mass'' (as for special relativity and mechanics). 

A purely mathematical theory as proposed here, though with limited (but relentlessly growing) domain of applicability, will have the eternal validity of mathematical truth. It will be a theory on which natural sciences can firmly rely. Such kind of theory is what I consider to be the solution of  the Sixth Hilbert's Problem. 

I argue that a prototype example of such a mathematical theory is provided by the novel algorithmic paradigm for physics, as in the recent information-theoretical derivation of quantum theory and free quantum field theory.
\end{abstract}
%%%%%%%%%%%%%%%%%%%%%%%%%%%

%%%%%%%%%% Insert the texts which can accomdate on firstpage in the tag "fmtext" %%%%%
\begin{fmtext}
The present opinion paper should be regarded mainly as a manifesto, preliminary to future  thorough studies.
\end{fmtext}

%%%%%%%%%%%%%%% End of first page %%%%%%%%%%%%%%%%%%%%%

\maketitle

\section{Mathematics is eternal, physics is temporary}

Who does not believe that mathematics is eternal? Will the Pitagora's theorem  continue to hold for the next millennia?

Differently from mathematical theorems, physical theories are temporary. Should they also be eternal? At first glance such possibility may look overreaching: physics has always been riven with contradictions, and contradictions have often played a pivotal role in the understanding of reality--a reality that, we believe, itself possesses internal logical coherence. But, is transience an intrinsic limitation of any physical theory? Or is it just a temporary historical feature? We cannot deny that, indeed, an impermanence phase occurred also for mathematics at its early pre-Hellenic stage, when the discipline was not so different from the ``trial-and error'' approach of physics (see the case of the value of Greek-Pi \cite{Eves}).

Physical theories are commonly considered eternal in the sense that, whenever the theory is replaced by a new one, %its validity 
nevertheless  the old theory continues to hold as a limiting case of the new theory, within a narrower phenomenological domain. Unfortunately this is not the case of the replacement of classical mechanics by quantum mechanics, since by no means the former can be regarded as a limiting case of the latter.\footnote{For example, no general procedure is known to recover the classical trajectories of a set of particles from the solution of the Schr\"odinger equation.} On the other hand, quantum mechanics, which supposedly holds for the entire physical domain, still relies on mechanical notions coming from the old classical theory, with the new theory built over the relics of the old one through heuristic {\em quantisation rules},\footnote{The canonical quantisation procedure is an extrapolation of the Ehrenfest theorem. For general functions over the phase-space one has  the well known issue of operator ordering, and it is only coincidental that Hamiltonians in nature are not affected by this problem. On curved space-time there exist no definite quantisation procedure. The geometric quantisation program has attempted to solve this problem, however, it never ended up with a general procedure. To date, it has succeeded only in unifying older methods of quantising finite-dimensional physical systems, but with no rule for infinite-dimensional systems, e.g. field theories and generally non integrable quantum systems (for a short and pedagogical review of the geometric quantisation, see Ref. \cite{Baez}).} or through the mathematically undefined {\em path integral} of Feynman.\footnote{The path integral works only as a formal rule for producing Feynman-diagrams expansions.} Issues of such extent definitely undermine the full logical coherence of the physical theory.

\section{The Sixth Hilbert Problem}

Was transience of physics what motivated David Hilbert to propose his Sixth Problem? We can imagine a positive answer from a mathematician who likely believed in permanence of mathematical truth.  Let's examine the opening paragraph of the problem. It reads: 
\\\par {\em The investigations on the foundations of geometry suggest the problem: To treat in the same manner by means of axioms, those physical sciences in which mathematics plays an important part; in the first rank are the theory of probabilities and mechanics.} 
\\ \par Benjamin Yandell writes in his book about Hilbert's problems \cite{Yandell}: 
\\ \par {\em Axiomatizing the theory of probabilities was a realistic goal: Kolmogorov accomplished this in 1933. The word 'mechanics' without a qualifier, however, is a Trojan horse.}
\\\par I argue that the Greek soldiers hiding inside the Trojan horse--and coming out of it much later in the fully developed theory--are the physical primitives contained in the axioms, i.e. in the same laws of mechanics. In fact, 
%The simple reason is that, 
being physical, the primitives require either a theoretical or an operational definition. This means that, in both cases, they are not actually primitive, implying that the logical feedback from their definition toward the same theory is prone to be logically not coherent.
 
Let's analyse the case of mechanics. ``Mechanics'' means dealing with ``forces'' and their effects on ``bodies''. Both notions of ``force'' and ``body'' are pregnant with theory, and are not actually primitive. For example, in Newtonian mechanics one needs to establish if a force is ``real'' or ``apparent'' (or, equivalently, whether the reference frame is ``inertial'' or not.\footnote{For Newton this was to a problem, since he believed in absolute space.}) However, only the theory can tell if a force is real (i.e. electromagnetic, weak, strong, ...), and this feeds the theory back into the axiomatic primitive in a circular way.\footnote{Notice that the notion of inertial frame is itself circular if nothing assesses reality of forces. One can introduce forces operationally through the Hooke law which, however, far from being primitive, is also operationally not defined at the microscopic level of particle physics.} On the other hand, the notion of ``body'', or physical ``object'' (on which forces act), is a seemingly innocent concept, which, however, bears its own theoretical logical discord. Indeed, the most general notion of object--the object as a ``bundle of properties''--is incompatible with quantum theory, since it lacks its {\em mereological} connotation (i.e. composability of objects to make new objects). In fact, in quantum theory it is common that the properties of the whole are incompatible with any possible property of the parts \cite{FQXi13}.\footnote{In quantum theory a property is associated to a projector operator over a subspace of the Hilbert space of the system (which we identify with the object). Joint properties are projectors on the tensor products of the Hilbert spaces of the component systems. Entangled tensor projectors do not commute with any local projector: this means that such properties of the whole are incompatible with any property of the parts. Moreover, entangled projectors are typical among projectors on tensor spaces.} To make things worst, the same notion of  ``particle'' as a localizable entity does not survive quantum field theory \cite{Clifton}, and what is left of the particle concept is a purely mathematical notion: the particle is an irreducible representation of the Poincar\'e a group \cite{Weinberg}. 

Keeping  Newtonian mechanics aside, and entering Einstein's special relativity, we encounter physical primitives as ``clock'' and ``rigid rod'', apparently unharmful notions entering the theory axiomatisation ``operationally'', but which instead are pregnant of theory, e.g. in regards of precision of clocks pace and rods length, which are both established by a separate theory: quantum theory!\footnote{Einstein himself was well aware of the primitiveness issue about the two notions of ``clock'' and ``rigid rod''. He writes: {\em It is . . . clear that the solid body and the clock do not in the conceptual edifice of physics play the part of irreducible elements, but that of composite structures, which must not play any independent part in theoretical physics. But it is my conviction that in the present stage of development of theoretical physics these concepts must still be employed as independent concepts; for we are still far from possessing such certain knowledge of the theoretical principles of atomic structure as to be able to construct solid bodies and clocks theoretically from elementary concepts.} \cite{Einstein}. For a critical analysis of the notions of clock and rigid rod in the axiomatisation of special relativity, see Chapt. 7 of the book of Harvey R. Brown \cite{Brownbook} along with the analysis of A. Hagar \cite{Hagar}.}

%\section{Group Theory is exemplar in the mathematisation of physics}
As an additional argument in support to a full mathematisation program of the physical theory we have the non accidental fact that what endures time in a theory is mainly its mathematical component. Exemplar in such respect is group theory, which provides a perfect case of purely mathematical formulation that is ``physical'' in the interpretation. And exemplar has been the success of group theory in the grand-unification of forces, ranging from the creation of the notion of spin and isospin, up to the quark model, and in providing the already mentioned geometrical categorisation of the notion of elementary particle as an irreducible representation of the Poincar\'e group.\footnote{Remarkably, such notion of particle genuinely wears an ontological version of structural realism, with the naive physical ontology replaced by a mathematical structure, whereas the physical interpretation comes in terms of a relational structure of reality. This shows how the process of mathematisation of physics also provides a unique reconciliation between the realist and the structuralist views.} 

\section{Failure of previous mathematisation programs}
The route for the mathematisation of physics has been deeply explored by Streater and Wightman \cite{Streater} and 
Haag and Kastler \cite{Haagb} for quantum field theory. These axiomatizations, however, still hinge on mechanical notions from the old classical theory, as those of  ``particle'', ``mass'',  but also ``space'' and ``time''. On the opposite side, the mathematical framework is inherited from quantum theory, based on the von Neumann axiomatics, which notoriously lacks physical interpretation. 

Irreconcilable with quantum field theory, the other main theory in physics--general relativity--uses space-time as a dynamical entity, whereas in quantum field theory it plays the role of a background required axiomatically. The  same ``equivalence principle'', which comes from experience, should originate from a microscopic mechanism giving rise to mass, but, instead, it is simply postulated.  On the other hand, the particle mass, a parameter which is unbounded in quantum field theory, becomes bounded as the result of a patch with general relativity, its maximum value, the Planck mass, corresponding to the notion of mini black-hole.\footnote{The Planck mass is the value of the mass for which the particle localisation size (the Compton wavelength) becomes comparable to the Schwarzild radius.} In the same way, patching gravity with quantum field theory clashes with the continuum description, as a consequence of the break down of the high-energy/short-distance correspondence, since colliding two particles with center-of-mass energy of the order of the Planck energy creates a black hole. This clash with general relativity must be summed to the logical mismatch entirely internal to quantum field theory: the Haag's theorem \cite{Streater,Haagb}.\footnote{Streater and Wightman write \cite{Streater}: {\em Haag's theorem is very inconvenient; it means that the interaction picture exists only if there is no interaction.} Haag's theorem states that there can be no interaction picture, that we cannot use the Fock space of noninteracting particles as a Hilbert space, in the sense that we would identify Hilbert spaces via field polynomials acting on a vacuum at a certain time. While some physicists and philosophers of physics have repeatedly emphasized how seriously Haag's theorem is shaking the foundations of the theory, practitioners simply dismiss the issue, whereas most quantum field theory texts do not even mention it.}

\section{Physics as interpretation of mathematics: the pet theory}
%The full mathematisation of physics}

The reader will ask again how is it possible to axiomatise a physical theory without resorting to physical notions. Physics without physics? Is it not an oxymoron? As already said, the answer is simply to restricts physics to the interpretational level. Then, if the interpretation is thorough for axioms, it will logically propagate throughout all theorems, up to the very last stage of their specific instances, where physics will emerge from the mathematics through its capability of portraying phenomena and to logically connect experimental observations. 

One may now legitimately ask from where can we learn the physics for the interpretation of the mathematics? A concrete possibility is the direct comparison of observations with the solutions that the mathematical algorithm portrays,  likewise a digital screen reproduces a physical scenario. Also, we can compare instances of the mathematical theory with our previous heuristic theoretical knowledge coming from non mathematically axiomatised theories. In the following I will provide a concrete example of how the ``mechanical'' quantum field theory can emerge in its customary differential form from a purely mathematical theory, as a special regime of a mathematical algorithm: a quantum walk.

The great bonus of having a mathematical theory endowed with physical interpretation, from axioms to theorems, will be the availability of thorough conceptual deductions that are physically meaningful at each logical step. The more universal the axioms are, the more stable and general the theory will be. And, contrarily to the case of derivations without physical interpretation (as for the quantum Hilbert space formalism), reasoning with mathematics with physical interpretation builds up a logically grounded physical insight.

\subsection{Quantum Theory as an information theory}
In order to be purely mathematical, the axioms cannot contain words requiring a physical definition, but must be meaningful for describing a physical scenario. How can this work?

An example of a purely mathematical framework with physical interpretation is probability theory. Here the notion of  ``event'' in its specific instances can contain everything is needed for the physical interpretation. And, indeed, probability theory is regularly used in physics with no interpretation issue. And, as already said, it has been axiomatised by Kolmogorov in 1933, thus solving the easiest part of the VI Hilbert problem.  

We want now to consider the case of quantum theory. With the term ``quantum theory''--instead of  ``quantum mechanics''--we denote the general theory of abstract systems, stripped of its mechanical connotation. The example of probability theory suggests rethinking the  theory as being itself an extension of probability theory, with the physical interpretation contained in the event specification. Von Neumann in his quantum-logic program \cite{Birvon,Hilvon} was seeking a similar route: indeed, the calculus of propositions is a purely mathematical context, with the interpretation contained in the specific proposition. Unfortunately, however, this program didn't reach the desired result. 

Instead of an ``alternative'' kind of logic, quantum theory can be axiomatised as an ``extension'' of logic. This is the case of the so-called operational probabilistic theory (OPT) \cite{CDP11,CUPbook,Hardy13}. An OPT is an extension of probability theory, to which one adds connectivity among events, in terms of a directed acyclic graph of input-output relations--the so-called ``systems'' of the theory. Mathematically this framework corresponds to a combination of category theory \cite{Coecke} with probability theory, and constitutes the mathematical backbone of a general ``theory of information'' \cite{CDPinCbook}. Therefore, an OPT is an extension of  probability theory, which in turn is an extension of logic (see Jaynes \cite{Jaynes} and Cox \cite{Cox}), and, in its spirit, this axiomatisation program remains close to the  quantum-logic one, which was missing the crucial ingredient of connectivity among events. (A concise complete exposition of the OPT framework can be found in Ref. \cite{CDPinCbook}.) 

Within the OPT framework, the six postulates of quantum theory are:
P1 {\em Atomicity of Composition}: The sequence of two atomic operations is atomic;\footnote{The ``operation'' is an event connected to others via systems. Here the term {\em atomic} means that the operation as event is not refinable into more elementary events.} 
P2 {\em Perfect Discriminability}: Every deterministic state that is not completely mixed is perfectly discriminable from some other state;
P3 {\em Ideal Compression}: Every state can be compressed in a lossless and efficient way;
P4 {\em Causality}: The probability of a preparation is independent of the choice of observation;
P5 {\em Local Discriminability}: It is possible to discriminate any pair of states of composite systems using only local observations;
P6 {\em Purification}: Every state can be always purified with a suitable ancilla, and two different purifications with the same ancilla are connected by a reversible transformation. (For a thorough analysis of the postulates and a didactical complete derivation of quantum theory, see the textbook \cite{CUPbook}: noticeably all the six postulates possess an epistemological connotation, concerning the falsifiability of propositions of the theory under local observations and control.) 

%Ultimately, the motivation for introducing physical notions in the axioms of a theory is the wish of connecting it to physical ontologies. In a structuralist point of view what matters are only the relations among the alleged ontologies, with theory being only an homomorphism of reality. 
The potential of the purely mathematical axiomatisation as an information theory is that physics can emerge from the network of  connections among events. The mechanical paradigm is substituted by the new algorithmic paradigm. The new paradigm brings a far reaching methodology change, since formulating the physical law as an  algorithm forces the theory to be expressed in precise mathematical terms. And the algorithm, regarded as a network of  ``operations'' bears also an interpretation as ``physical algorithm'', e.g. an experimental protocol.
%This is the case of the six information-theoretic postulates of quantum theory \cite{CDP11}, mathematically defined within the framework of strict symmetric monoidal category theory\cite{CDPinCbook}, and having physical interpretation in terms of ``physical algorithms'', namely experimental protocols.
Quantum theory is thus a special kind of information theory, namely a theory of {\em processes}, 
%--relations without ontologies\footnote{It should be noted that the same can be said of classical information theory, which indeed share five postulates with quantum theory. The distinctive axiom of quantum theory is the so called ``purification axiom'', which provides maximal control of randomness \cite{CDP11}.}--
describing a network of connections among events, some of which are under our control (the so-called {\em preparations} or {\em states}),  others are what we observe (the so-called {\em observations} or {\em effects}), some others are {\em transformations}, connecting preparations with observations.\footnote{It should be noted that the same can be said of classical information theory, which indeed share five postulates with quantum theory. The distinctive axiom of quantum theory is the purification postulate P6, which provides maximal control of randomness \cite{CUPbook}.}  

\subsection{How ``mechanics'' emerges from information: free quantum field theory}
All what was said above pertains only to the quantum theory of abstract systems. But, how can the ``mechanics'', namely quantum field theory, emerge? The answer resides in adding rules to the process in terms of general topological requirements for the connectivity network, where the connectivity is given by systems interactions. Such requirements are (informally): P7 {\em Unitarity}: the evolution is unitary; P8 {\em Locality}: each system interacts with a uniformly bounded number of systems; P9 {\em Homogeneity}: the network of interactions is indistinguishable when regarded from any two quantum systems; P10 {\em Isotropy} (see \cite{DPDirac}). All postulates P7-P10 correspond to minimise the algorithmic complexity of the quantum process.\footnote{Since every deterministic transformation can be achieved unitarily in quantum theory, P7 represents the choice of minimal complexity. P8 and P9 correspond to infinite reduction of complexity, and P10 is also a reduction of complexity. The complexity of  the quantum cellular automaton can be quantified in terms of the complexity of a minimal Margolus quantum circuit-block \cite{Toffoli-Margolus,Margolus} for the automaton, by counting the resources of the circuit in terms of controlled-not and Pauli unitaries according to Deutsch theorem \cite{Deutsch}. Differently from the Kolmogorov-Chaitin complexity \cite{Kolm1,Kolm2}, the complexity of the automaton is computable, and it is easy to  get a very good upper bound. This will then quantify the complexity of the physical law, toward a quantification of the Occam's razor criterion.} Postulate P9 requires that the graph of interactions is the Cayley graph of a finitely presented group. Then, for the easiest case of: a) Abelian group, b) linearity of the evolution, and c) minimal dimension $s$ of the system for nontrivial evolution ($s=2$) one finds a quantum walk which in the limit of small wave vectors gives the Weyl equation \cite{DPDirac}. For $s>2$ the solution of the unitarity conditions becomes increasingly hard:  two solutions can be easily provided for $s=4$ in terms of the direct sum and the tensor product of  two copies of the quantum walk with $s=2$: these correspond to Dirac \cite{DPDirac} and Maxwell \cite {BDPMaxwell} field theories, respectively. Noticeably the relativistic quantum field theory is obtained without using any mechanics, any relativistic covariance, and not even space-time.\footnote{Lorentz covariance is a result of the Galileo principle, along with causality, locality, homogeneity, and isotropy of interactions of abstract quantum systems. The inertial frames are those that leave the quantum automaton or quantum walk invariant\cite{Lorentz}.} And being quantum {\em ab initio}, the theory does not need quantisation rules. (For a review, the reader is addressed to Ref. \cite{IJTP}).  The ``mechanics''--including space-time and Lorentz covariance emerges from a purely mathematical algorithm, at the large scale of a theory that is inherently discrete. Physics emerges in form of computational patterns, in a universe which resembles an  immense quantum-digital screen.

\subsection{How physical standards come out in a purely mathematical theory}
In a mathematically formulated theory all variables with physical interpretation are necessarily adimensional. Therefore, the physical interpretation of the theory will be complete only if the same theory contains a way to establish the physical standards. In a purely mathematical theory this can be done either through counting, or by comparing real numbers with special values of the theory--typically minimal or maximal allowed values of adimensional variables pregnant with physical interpretation. The quantum cellular automata theory of fields of the previous section \cite{DPDirac} is exemplar in this sense, as it sets measurements of space and time to counting, whereas the interplay of unitarity and discreteness leads to an upper bound for the particle mass \cite{DPDirac}, which can then plays the role of the mass standard.\footnote{Since at the maximum value of the mass there is no information flow in the quantum walk, the same value is interpreted as that of the mini black-hole originating from the patching of quantum field theory with general relativity. Thus the mass standard is the already mentioned Planck mass. Moreover, from the comparison with the usual free quantum field theory emerging at large scales (small wave vectors) one finds that the discrete units of time and space are the Planck's units.}

\section{Conclusions}
Is mathematisation of physics still premature? Although we may never achieve a complete set of axioms for physics, yet we can have stable subsets of them (along with alternative equivalent sets), and derive a stable collection of theorems, whose physical interpretation will provide us with ``physical'' truth as evident and eternal as the mathematical one. Mathematics is an evolution of the human language--the ultimate idealisation and synthesis of observations grouped into equivalence classes. Physics goes beyond logic, and, as such, it is an interpretation of mathematics in terms of our experience. A thorough mathematical axiomatisation program for physics will open a new era, purely logical, not conjectural, focused on seeking general principles whose resulting theory will retain in time autonomous mathematical value. 

Four hundred years already passed after Galileo introduced the mathematical description of physics: the complete mathematisation of physics will be the next, ultimate, Galilean revolution. 
\bigskip
\disclaimer{The fact that the author calls for a mathematisation of physics does not imply that he is a good mathematician. 
Reference to author's pet theories is only for the sake of exemplification, although these theories were designed and elaborated exactly within the spirit and motivations of the present essay. The reader should judge the present mathematisation program independently on his opinions on these theories.}
\dataccess{For a recent review about the author's pet theory here mentioned, see the recent open access review by the same author in memoriam of his friend David Finkelstein \cite{IJTP} \doi{10.1007/s10773-016-3172-y}}

\funding{This work was made possible through the support of a grant from the John Templeton Foundation (Grant ID\# 43796: {\em A Quantum-Digital Universe}, and Grant ID\# 60609: {\em Quantum Causal Structures}). The opinions expressed in this publication are those of the author and do not necessarily reflect the views of the John Templeton Foundation. The manuscript has been mostly written at the Kavli Institute for Theoretical Physics, University of California, Santa Barbara, and supported in part by the National Science Foundation under Grant No. NSF PHY11-25915.} 
\ack{This paper is a largely updated version of the original essay: {\em The unreasonable effectiveness of Mathematics in Physics: the Sixth Hilbert Problem,  and the Ultimate Galilean Revolution}, written for the FQXi Contest 2015: {\em Trick or Truth}, 
\url{http://fqxi.org/community/forum/topic/2395}.}

\end{document}